\documentclass[aps,preprint,tightenlines,superscriptaddress]{revtex4}%
\usepackage{amsfonts}
\usepackage{amsmath}
\usepackage{amssymb}
\usepackage{graphicx}%
\providecommand{\U}[1]{\protect\rule{.1in}{.1in}}

\begin{document}
\preprint{ANL-HEP-PR-09-14 and UMTG-9}
\title[C\&QTAs]{Classical and Quantal Ternary Algebras}
\author{Thomas Curtright}
\affiliation{Department of Physics, University of Miami, Coral Gables, FL 33124-8046, USA}
\author{David Fairlie}
\affiliation{Department of Mathematical Sciences, Durham University, Durham, DH1 3LE, UK}
\author{Xiang Jin}
\affiliation{Department of Physics, University of Miami, Coral Gables, FL 33124-8046, USA}
\author{Luca Mezincescu}
\affiliation{Department of Physics, University of Miami, Coral Gables, FL 33124-8046, USA}
\author{Cosmas Zachos}
\affiliation{High Energy Physics Division, Argonne National Laboratory, Argonne, IL
60439-4815, USA\medskip\medskip}
\keywords{one two three}
\pacs{PACS number}

\begin{abstract}
\medskip

We consider several ternary algebras relevant to physics. \ We compare and
contrast the quantal versions of the algebras, as realized through associative
products of operators, with their classical counterparts, as realized through
classical Nambu brackets. \ In some cases involving infinite algebras, we show
the classical limit may be obtained by a contraction of the quantal algebra,
and then explicitly realized through classical brackets. \ We illustrate this
classical-contraction method by the Virasoro-Witt example.

\end{abstract}
\volumeyear{year}
\volumenumber{number}
\issuenumber{number}
\eid{identifier}
\startpage{1}
\endpage{ }
\maketitle

\section{Introduction}

Ternary algebras can be realized in two ways: \ They may be built from
antisymmetrized products of three linear operators -- so-called quantal
brackets -- or they may be realized through a generalization of Poisson
brackets in terms of multivariable Jacobians -- so-called classical brackets.
\ Both constructions were introduced in the physics literature by Nambu
\cite{N} after about twenty years of gestation. \ Lately, these algebras seem
to be gaining in usefulness and importance to physics. \ For example, there
has been some progress in constructing a $2+1$-dimensional local quantum field
theory with $SO(8)$ superconformal symmetry \cite{B,S,G,Gustavsson,H} as a
stepping-stone to obtain a world-volume Lagrangian description for coincident
M2-branes. \ The use of \ ternary algebras is crucial to the construction. \ 

More generally, \emph{N-algebras} involve antisymmetric operations on N
entities at a time. The mathematics was developed by Filippov in a paper that
appeared just over a decade after Nambu's \cite{F}, although Filippov does
\emph{not} seem to have been aware of Nambu's earlier work
\cite{CFZ08Misstatement}. \ In any case, both Nambu's and Filippov's work
motivated and inspired other mathematical investigations of these ideas
\cite{Bremner1995,Gautheron,T,Vaisman} along with many other physics studies
(see \cite{CZ2002,CZ} and references therein, especially \cite{deA,Hoppe}), as
well as much more recent work
\cite{Ahn,Axenides,Bonelli,Chu,Garousi,Hanaki,Larsson,KM}. \ \ 

Here we compare classical and quantal realizations of ternary algebras for
several interesting cases relevant to physics. \ The quantal and classical
algebras are usually \emph{not} the same, but for one very special case they
are: \ Nambu's $su\left(  2\right)  $. \ For some infinite cases, especially
the Virasoro-Witt algebra \cite{CFZ08}, the classical limit may be obtained
initially by a contraction of the quantal algebra, and then explicitly
realized through classical 3-brackets. \ 

\section{Classical and quantal brackets}

In this section, we define and compare the properties of various realizations
of 3-brackets. \ These are just\ trilinear operations performed either on
three operators or on three functions defined on a manifold. \ For the latter
case, some further discussion of the manifold's classical geometric structure
is given in an Appendix.

\subsection{Three specific realizations of 3-brackets}

The classical Nambu bracket \cite{N} based on three variables (say $x,y,z$) is
the simplest to compute, in most situations. \ It involves the Jacobian-like
determinant of partial derivatives of three functions $A,B,C$.
\begin{equation}
\left\{  A,B,C\right\}  =\frac{\partial\left(  A,B,C\right)  }{\partial\left(
x,y,z\right)  }=\varepsilon^{abc}~\partial_{a}A~\partial_{b}B~\partial_{c}C\ ,
\label{Nambu3CNB}%
\end{equation}
where $a,b,c$ are implicitly summed over $1,2,3$. \ The classical bracket is
totally antisymmetric in the three argument functions. \ (The $\left\{
,\right\}  $ notation is here used to distinguish this classical case from the
quantal bracket, and should not be confused with anti-commutation.)

The operator or \textquotedblleft quantal\textquotedblright\ 3-bracket was
originally defined \cite{N} to be a totally antisymmetrized sum of
trinomials,
\begin{equation}
\left[  A,B,C\right]  =ABC-BAC+CAB-ACB+BCA-CBA\ . \label{3QNB}%
\end{equation}
regardless of the number of underlying independent variables. \ This can be
equivalently expressed as a sum of single operators producted with commutators
of the remaining two, or as anticommutators acting on the commutators,
\begin{equation}
\left[  A,B,C\right]  =A\left[  B,C\right]  +B\left[  C,A\right]  +C\left[
A,B\right]  =\left[  B,C\right]  A+\left[  C,A\right]  B+\left[  A,B\right]
C\ . \label{Rewritten3QNB}%
\end{equation}
In principle, as well as in practice, it is necessary to have some
\emph{information about products}, hence about anticommutators as well as
commutators, to perform the actual evaluation of this quantal bracket.
\ Relatedly, the trace of $\left[  A,B,C\right]  $\ is non-trivial, in general.

There is yet another definition of an operator 3-bracket, introduced by Awata,
Li, Minic, and Yoneya \cite{A}, which is valid for operators of trace class.
It consists of re-packaging the commutators of any Lie algebra to define
\begin{equation}
\left\langle A,B,C\right\rangle =\left[  A,B\right]  ~Tr\left(  C\right)
+\left[  C,A\right]  ~Tr\left(  B\right)  +\left[  B,C\right]  ~Tr\left(
A\right)  \ . \label{ALMY}%
\end{equation}
This is again a totally antisymmetric trilinear combination, but it is a
singular construction for finite dimensional realizations in the sense that
$Tr\left\langle A,B,C\right\rangle =0$. \ For reasons to be discussed, this
ALMY bracket has properties intermediate between the classical 3-bracket and
the full quantal bracket.

\subsection{Properties of the various brackets}

As mentioned already, all three are trilinear and totally antisymmetric. \ On
the other hand, only two out of three \emph{automatically} satisfy the
Filippov condition\ \cite{F}, the so-called \textquotedblleft
FI\textquotedblright\ (also see \cite{Sahoo,T}). \ Namely,%
\begin{align}
\left\{  A,B,\left\{  C,D,E\right\}  \right\}   &  =\left\{  \left\{
A,B,C\right\}  ,D,E\right\}  +\left\{  C,\left\{  A,B,D\right\}  ,E\right\}
+\left\{  C,D,\left\{  A,B,E\right\}  \right\}  \ ,\nonumber\\
\left\langle A,B,\left\langle C,D,E\right\rangle \right\rangle  &
=\left\langle \left\langle A,B,C\right\rangle ,D,E\right\rangle +\left\langle
C,\left\langle A,B,D\right\rangle ,E\right\rangle +\left\langle
C,D,\left\langle A,B,E\right\rangle \right\rangle \ .
\end{align}
So, for the classical and ALMY brackets, the FI is indeed an identity, hence
it is a \emph{necessary} condition to realize a ternary algebra in terms of
either of these 3-brackets. \ But in general the FI does not hold for
associative operator products.%
\begin{align}
0  &  \neq\left[  A,B,\left[  C,D,E\right]  \right]  -\left[  \left[
A,B,C\right]  ,D,E\right]  -\left[  C,\left[  A,B,D\right]  ,E\right]
-\left[  C,D,\left[  A,B,E\right]  \right] \nonumber\\
&  \equiv\mathrm{fi}\left(  A,B;C,D,E\right)  \ .
\end{align}
That is to say, the Filippov condition, $\mathrm{fi}\left(  A,B;C,D,E\right)
=0$, is \emph{not} an operator identity. \ It holds only in special
circumstances. \ In this sense the FI differs from the Jacobi identity for
associative operator products, a \emph{2-bracket-acting-on-2-bracket}
situation. \ Looking ahead, we write the Jacobi identity in the somewhat
unusual form%
\begin{equation}
2\varepsilon^{ij}\left[  \left[  A,B_{i}\right]  ,B_{j}\right]  =\varepsilon
^{ij}\left[  A,\left[  B_{i},B_{j}\right]  \right]  \ ,
\end{equation}
where\ $A$ is fixed but all indexed entries are implicitly summed over
$i,j=1,2$, hence antisymmetrized.

Rather remarkably, all three types of 3-brackets \emph{do} satisfy an identity
first found by Bremner (henceforth, the \textquotedblleft BI\textquotedblright%
)\textbf{, }a \emph{3-on-3-on-3} multiple bracket relation
\cite{Bremner,Nuyts}.
\begin{align}
\varepsilon^{i_{1}\cdots i_{6}}\left\{  \left\{  A,\left\{  B_{i_{1}}%
,B_{i_{2}},B_{i_{3}}\right\}  ,B_{i_{4}}\right\}  ,B_{i_{5}},B_{i_{6}%
}\right\}   &  =\varepsilon^{i_{1}\cdots i_{6}}\left\{  \left\{  A,B_{i_{1}%
},B_{i_{2}}\right\}  ,\left\{  B_{i_{3}},B_{i_{4}},B_{i_{5}}\right\}
,B_{i_{6}}\right\}  \ ,\nonumber\\
\varepsilon^{i_{1}\cdots i_{6}}\left\langle \left\langle A,\left\langle
B_{i_{1}},B_{i_{2}},B_{i_{3}}\right\rangle ,B_{i_{4}}\right\rangle ,B_{i_{5}%
},B_{i_{6}}\right\rangle  &  =\varepsilon^{i_{1}\cdots i_{6}}\left\langle
\left\langle A,B_{i_{1}},B_{i_{2}}\right\rangle ,\left\langle B_{i_{3}%
},B_{i_{4}},B_{i_{5}}\right\rangle ,B_{i_{6}}\right\rangle \ ,\nonumber\\
\varepsilon^{i_{1}\cdots i_{6}}\left[  \left[  A,\left[  B_{i_{1}},B_{i_{2}%
},B_{i_{3}}\right]  ,B_{i_{4}}\right]  ,B_{i_{5}},B_{i_{6}}\right]   &
=\varepsilon^{i_{1}\cdots i_{6}}\left[  \left[  A,B_{i_{1}},B_{i_{2}}\right]
,\left[  B_{i_{3}},B_{i_{4}},B_{i_{5}}\right]  ,B_{i_{6}}\right]  \ .
\end{align}
where again, $i_{1},\cdots,i_{6}$ are implicitly summed from $1$ to $6$. \ For
the classical bracket in (\ref{Nambu3CNB}), both left- and right-hand-sides of
the BI actually vanish. \ For the ALMY bracket, the identity follows by direct calculation.

For quantal 3-brackets the BI is \emph{a consequence of associativity}. \ If
one posits an operator 3-bracket based on associative products, and it does
not satisfy this identity, then one has \emph{erred}. \ Thus the BI is a
\emph{necessary} condition to realize a ternary algebra in terms of operator
brackets. \ 

Finally, we note that the antisymmetrization of an operator 3-bracket acting
on a second 3-bracket does \emph{not} close to give a third 3-bracket, but
rather a \emph{5-bracket}: \ $\varepsilon^{i_{1}\cdots i_{5}}\left[  \left[
B_{i_{1}},B_{i_{2}},B_{i_{3}}\right]  ,B_{i_{4}},B_{i_{5}}\right]
\propto\left[  B_{1},B_{2},B_{3},B_{4},B_{5}\right]  $. \ The operator
5-bracket is defined in general \cite{CZ} as the totally antisymmetric signed
sum over all $5!$ distinct words $ABCDE$, etc.

\section{Examples of ternary algebras.}

When the 3-brackets close to yield other functions or operators of a
designated set, one is considering a ternary algebra. \ In this event, one may
define structure constants to write, say for the quantal bracket,%
\begin{equation}
\left[  A_{i},A_{j},A_{k}\right]  =f_{ijk}^{\ \ \ l}~A_{l}\ .
\end{equation}
Similarly for the classical and ALMY brackets. \ For the latter two types of
brackets, the FI always holds, and this implies a bilinear consistency
condition on the respective structure constants. \ This is similar to the
consistency condition imposed on the structure constants of a Lie algebra by
the Jacobi identity. \ However, this bilinear condition does not necessarily
apply to the quantal structure constants \cite{CZ}. \ 

More generally, the BI gives a trilinear consistency condition on the
structure constants for all three types of brackets. \ This trilinear
condition must always be satisfied by any ternary algebra. \ For the classical
and ALMY brackets, the BI structure constant condition is not independent of
the FI condition. \ But for a quantal 3-algebra the BI condition is the
\emph{only} constraint imposed on the $f_{ijk}^{\ \ \ l}$\ by associativity.
\ It would be interesting to classify all quantal 3-algebras by constructing
all solutions of the BI condition on the $f_{ijk}^{\ \ \ l}$. \ We leave this
for the well-motivated reader to pursue. \ Here, we just consider various
basic examples of intrinsic interest to physics.

\subsection{Nambu's $su(2)$}

First, consider\ Nambu's application to $su(2).$%
\begin{equation}
\left[  L_{x},L_{y},L_{z}\right]  \equiv L_{x}\left[  L_{y},L_{z}\right]
+L_{y}\left[  L_{z},L_{x}\right]  +L_{z}\left[  L_{x},L_{y}\right]  =i~\left(
L_{x}^{2}+L_{y}^{2}+L_{z}^{2}\right)  \ .
\end{equation}
To close the algebra, it is necessary to include the $su(2)$\ Casimir. \ But,
having done so, one may rescale by a fourth root of the Casimir%
\begin{equation}
Q_{x}=\frac{L_{x}}{\sqrt[4]{L^{2}}}\ ,\ \ \ Q_{y}=\frac{L_{y}}{\sqrt[4]{L^{2}%
}}\ ,\ \ \ Q_{z}=\frac{L_{z}}{\sqrt[4]{L^{2}}}\ ,
\end{equation}
and define a fourth charge as that fourth root,%
\begin{equation}
Q_{t}=\sqrt[4]{L^{2}}\ .
\end{equation}
Then,%
\begin{equation}
\left[  Q_{a},Q_{b},Q_{c}\right]  =i~\varepsilon_{abcd}~Q^{d}\ .
\label{NambuSU(2)}%
\end{equation}
where $\varepsilon_{xyzt}=+1$ with a $\left\lceil -1,-1,-1,+1\right\rfloor $
Lorentz signature. \ The usual $\varepsilon\varepsilon$ identities now imply
that this example is special: \ The FI holds for Nambu's $su\left(  2\right)
$. \ In fact, this is the \emph{only} quantal ternary algebra that satisfies
the FI, and, as we shall see, it has an explicit realization through classical
3-brackets\textit{.} \ 

Toward that end, we note that Nambu $su(2)$ has sub-3-algebras that close.
\ They are easily found. \ For example,%
\begin{equation}
Q_{x}\ ,\ \ \ Q_{y}\ ,\ \ \ Q_{z}\pm Q_{t}\ .
\end{equation}
Moreover, each of these subalgebras can be realized in terms of the 3-bracket
(\ref{Nambu3CNB}), as is evident if we just\ consider the brackets of%
\begin{equation}
x\sqrt{z}\ ,\ \ \ y\sqrt{z}\ ,\ \ \ z\text{ \ \ and \ \ }x\sqrt{z}%
\ ,\ \ \ y\sqrt{z}\ ,\ \ \ x^{2}+y^{2}\ .
\end{equation}
Thus the complete 3-algebra of all four $Q_{a}$\ can be realized in terms of
classical brackets. \ Having done so, it follows without calculation that the
structure constants of this 3-algebra are such that the FI is satisfied
identically, since the Nambu 3-brackets in (\ref{Nambu3CNB}) always obey this
condition. \ In contrast, however, we note that (\ref{NambuSU(2)}) cannot be
realized as ALMY brackets, since $\sqrt[4]{L^{2}}=Q_{t}=-i\left[  Q_{x}%
,Q_{y},Q_{z}\right]  $ is \emph{not} traceless.

\subsection{The bosonic oscillator. \ }

The usual four charges $1,\ a,\ a^{\dagger},$ and$\ N=a^{\dagger}a$\ give the
quantal ternary algebra%
\begin{equation}
\left[  1,N,a\right]  =-a\ ,\ \ \ \left[  1,N,a^{\dagger}\right]  =a^{\dagger
}\ ,\ \ \ \left[  1,a,a^{\dagger}\right]  =1\ ,\ \ \ \left[  N,a,a^{\dagger
}\right]  =-1-N\ . \label{Oscillator}%
\end{equation}
Three of these reduce to just commutators: \ $\left[  1,N,a\right]  =\left[
N,a\right]  ,\ \left[  1,N,a^{\dagger}\right]  =\left[  N,a^{\dagger}\right]
,$ and$\ \left[  1,a,a^{\dagger}\right]  =\left[  a,a^{\dagger}\right]  $.
\ This would suggest that the same algebra might also be realized as
ALMY\ brackets, except for the fact that the operators at hand are \emph{not}
of trace class. \ In any case, the fourth relation in (\ref{Oscillator}) is
not so simple.

However, if we take linear combinations as%
\begin{equation}
R_{1}=N\ ,\ \ \ R_{2}=\frac{1}{\sqrt{2}}\left(  a^{\dagger}+a\right)
\ ,\ \ \ R_{3}=\frac{1}{\sqrt{2}}i\left(  a^{\dagger}-a\right)  \ ,\ \ \ R_{4}%
=N+1\ ,
\end{equation}
then we are back to a variation on Nambu's theme for $su\left(  2\right)  $:
\ In this case, $sl\left(  2,%
\mathbb{R}
\right)  $.%
\begin{equation}
\left[  R_{a},R_{b},R_{c}\right]  =i\epsilon_{abcd}~R^{d}\ ,
\end{equation}
with $\epsilon_{1234}=+1$, again with Lorentz metric to raise indices,
$\eta_{ab}=\left\lceil 1,1,1,-1\right\rfloor $. \ So, what's new here? \ 

There are two additional bilinears, $a^{2}$ and $a^{\dagger2}$, whose
3-brackets give oscillator trilinears.%
\begin{align}
\left[  a,a^{2},a^{\dagger2}\right]   &  =2a+2Na\ ,\ \ \ \left[  a^{\dagger
},N,a^{2}\right]  =-2a-Na\ ,\ \ \ \left[  a,N,a^{2}\right]  =-a^{3}%
\ ,\nonumber\\
\left[  a^{\dagger},a^{2},a^{\dagger2}\right]   &  =2a^{\dagger}+2a^{\dagger
}N\ ,\ \ \ \left[  a,N,a^{\dagger2}\right]  =2a^{\dagger}+a^{\dagger
}N\ ,\ \ \ \left[  a^{\dagger},N,a^{\dagger2}\right]  =a^{\dagger3}\ .
\end{align}
Therefore, upon closure, the ternary algebra becomes \emph{infinite}, and the
standard enveloping algebra for the oscillator is obtained. \ From Filippov's
perspective, it is perhaps disappointing that the oscillator enveloping
algebra does not satisfy the FI. \ For example%
\begin{align}
-2  &  =\left[  \left[  a^{\dagger},a^{\dagger}a,a^{\dagger2}\right]
,a,a^{2}\right]  -\left[  \left[  a^{\dagger},a,a^{2}\right]  ,a^{\dagger
}a,a^{\dagger2}\right]  -\left[  a^{\dagger},\left[  a^{\dagger}%
a,a,a^{2}\right]  ,a^{\dagger2}\right]  -\left[  a^{\dagger},a^{\dagger
}a,\left[  a^{\dagger2},a,a^{2}\right]  \right]  \ ,\nonumber\\
20a^{\dagger}  &  =\left[  \left[  a^{\dagger}a,a^{\dagger2},a^{2}\right]
,a,a^{\dagger2}\right]  -\left[  \left[  a^{\dagger}a,a,a^{\dagger2}\right]
,a^{\dagger2},a^{2}\right]  -\left[  a^{\dagger}a,\left[  a^{\dagger
2},a,a^{\dagger2}\right]  ,a^{2}\right]  -\left[  a^{\dagger}a,a^{\dagger
2},\left[  a^{2},a,a^{\dagger2}\right]  \right]  \ .
\end{align}
In any case, the FI does \emph{not} hold in this example. \ But, necessarily,
the associative enveloping algebra \emph{does} satisfy the BI.

\subsection{Virasoro--Witt 3-algebra.}

For the oscillator there is a familiar, infinite Lie algebra contained within
the enveloping algebra. \ Consider
\begin{equation}%
\mathcal{L}%
_{n}=-\left(  a^{\dagger}\right)  ^{n}N\ , \label{SimpleVirasoro}%
\end{equation}
for $n\geq0$. \ Commutators give the well-known Virasoro-Witt (VW) algebra.%
\begin{equation}
\left[
\mathcal{L}%
_{n},%
\mathcal{L}%
_{m}\right]  =\left(  n-m\right)
\mathcal{L}%
_{n+m}\ ,
\end{equation}
for $m,n\geq0$. \ It is less well-known that the corresponding quantal
3-brackets are%
\begin{equation}
\left[
\mathcal{L}%
_{n},%
\mathcal{L}%
_{m},%
\mathcal{L}%
_{k}\right]  =0\ .
\end{equation}
Thus we have a \emph{null 3-algebra} for an infinite set of non-trivial,
non-commuting oscillator charges. \ The FI is trivially satisfied in this
case, as is the BI.

More structure is evident if we slightly modify the oscillator realization of
the VW algebra. \ For parameters $\beta$ and $\gamma$, define and compute,%
\begin{equation}
L_{n}=-\left(  a^{\dagger}\right)  ^{n}\left(  N+\gamma+n\beta\right)
\ ,\ \ \ \left[  L_{n},L_{m}\right]  =\left(  n-m\right)  L_{n+m}\ .
\label{LessSimpleVirasoro}%
\end{equation}
The parameter $\beta$ is related to the $sl\left(  2,%
\mathbb{R}
\right)  $ Casimir, $C=\beta\left(  1-\beta\right)  $. \ Now we find a
non-null quantal 3-bracket when $0\neq\beta\neq1$.%
\begin{equation}
\left[  L_{n},L_{m},L_{k}\right]  =\beta\left(  1-\beta\right)  \ \left(
n-m\right)  \left(  m-k\right)  \left(  n-k\right)  \ M_{n+m+k}\ ,
\label{3LVW}%
\end{equation}
where a second sequence of charges has been defined by
\begin{equation}
M_{n}=\left(  a^{\dagger}\right)  ^{n}\ . \label{VirasoroCompanion}%
\end{equation}
While the Lie algebra of the $L$s and $M$s is also well-known \cite{GT}, their
3-algebra has been investigated only recently \cite{CFZ08}. \ To close the
3-algebra, we must consider all additional 3-brackets involving the $M$s:%
\begin{align}
\left[  M_{n},L_{m},L_{k}\right]   &  =\left(  m-k\right)  \left(
L_{n+m+k}+\left(  1-2\beta\right)  nM_{n+m+k}\right)  \ ,\nonumber\\
\left[  M_{n},M_{m},L_{k}\right]   &  =\left(  m-n\right)  M_{n+m+k}%
\ ,\nonumber\\
\left[  M_{n},M_{m},M_{k}\right]   &  =0\ . \label{RestOf3VW}%
\end{align}
While the calculation is involved, the BI may be confirmed to hold for this
ternary algebra. \ This result follows from the use of only (\ref{3LVW}) and
(\ref{RestOf3VW}), and does not make explicit use of the oscillator
realization employed to obtain the 3-algebra. \ So this algebra is consistent
with an underlying associative operator product no matter how it is realized.

The modification of the oscillator realization to include the parameter
$\beta$ has led to a larger ternary algebra involving both $L$s and $M$s.
\ But, so enlarged, the algebra as presented in (\ref{3LVW}) and
(\ref{RestOf3VW}) is cumbersome. \ It may be streamlined by a linear change of
basis, effectively from $L_{n}$ and $M_{n}$ back to the original $%
\mathcal{L}%
_{n}$ and $M_{n}$, as in (\ref{SimpleVirasoro}) and (\ref{LessSimpleVirasoro}%
). \ That is to say, let
\begin{equation}%
\mathcal{L}%
_{n}\equiv L_{n}+\left(  \gamma+\beta n\right)  M_{n}\ .
\end{equation}
Regardless of how the algebra is realized, this change of basis simplifies
(\ref{3LVW}) and (\ref{RestOf3VW}). \ We find a remarkably concise form for
the ternary algebra.%
\begin{gather}
\left[
\mathcal{L}%
_{k},%
\mathcal{L}%
_{m},%
\mathcal{L}%
_{n}\right]  =0\ ,\nonumber\\
\left[  M_{k},M_{m},M_{n}\right]  =0\ ,\nonumber\\
\left[
\mathcal{L}%
_{k},M_{m},M_{n}\right]  =\left(  n-m\right)  M_{k+m+n}\ ,\nonumber\\
\left[  M_{k},%
\mathcal{L}%
_{m},%
\mathcal{L}%
_{n}\right]  =\left(  m-n\right)  \left(
\mathcal{L}%
_{k+m+n}-kM_{k+m+n}\right)  \ . \label{Simplified3Virasoro}%
\end{gather}
All explicit $\beta$ dependence is thereby removed from the 3-algebra in this
basis, for \emph{all} values of the $sl\left(  2,%
\mathbb{R}
\right)  $ Casimir. \ Now, what about the FIs? \ 

The FIs sometimes fail. \ This was discussed in \cite{CFZ08}, in the original
basis, but it is much more transparent in terms of (\ref{Simplified3Virasoro}%
). \ It is trivial to see that the Filippov condition is satisfied when only $%
\mathcal{L}%
$s, or when only $M$s, are involved: $\mathrm{fi}\left(
\mathcal{L}%
_{p},%
\mathcal{L}%
_{q};%
\mathcal{L}%
_{k},%
\mathcal{L}%
_{m},%
\mathcal{L}%
_{n}\right)  =0=\mathrm{fi}\left(  M_{p},M_{q};M_{k},M_{m},M_{n}\right)  $.
\ The condition is also satisfied when there are two, three, or four $M$s
mixing it up with $%
\mathcal{L}%
$s. \ But when one $M$\ is entangled with four $%
\mathcal{L}%
$s, the condition fails, in general:%
\begin{align}
\mathrm{fi}\left(
\mathcal{L}%
_{p},%
\mathcal{L}%
_{q};%
\mathcal{L}%
_{k},%
\mathcal{L}%
_{m},M_{n}\right)   &  =\left(  p-q\right)  \left(  k-m\right)  \left(
k+m-p-q\right)  n~M_{k+m+n+p+q}\ ,\nonumber\\
\mathrm{fi}\left(
\mathcal{L}%
_{p},M_{q};%
\mathcal{L}%
_{k},%
\mathcal{L}%
_{m},%
\mathcal{L}%
_{n}\right)   &  =\left(  n-k\right)  \left(  k-m\right)  \left(  m-n\right)
q~M_{k+m+n+p+q}\ .
\end{align}
On the other hand, the BI is again seen to always hold. \ We stress that these
results follow from the use of only (\ref{Simplified3Virasoro}) without
explicit use of the oscillator realization.

The situation with the FIs can be remedied if we perform an
In\"{o}n\"{u}-Wigner contraction \cite{IW}. \ This produces an algebra that
satisfies the FI in \emph{all} cases. \ The procedure is to rescale $%
\mathcal{L}%
_{k},~M_{k}\mapsto\mathfrak{L}_{k}\equiv\lambda^{-1}%
\mathcal{L}%
_{k},~\mathfrak{P}_{k}\equiv\lambda M_{k}$ and take the formal limit
$\lambda\rightarrow\infty$. \ The result is just to discard the term
$kM_{k+m+n}$ in the last line of (\ref{Simplified3Virasoro}). \
\begin{gather}
\left[  \mathfrak{L}_{k},\mathfrak{L}_{m},\mathfrak{L}_{n}\right]
=0\ ,\nonumber\\
\left[  \mathfrak{P}_{k},\mathfrak{P}_{m},\mathfrak{P}_{n}\right]
=0\ ,\nonumber\\
\left[  \mathfrak{L}_{k},\mathfrak{P}_{m},\mathfrak{P}_{n}\right]  =\left(
n-m\right)  \mathfrak{P}_{k+m+n}\ ,\nonumber\\
\left[  \mathfrak{P}_{k},\mathfrak{L}_{m},\mathfrak{L}_{n}\right]  =\left(
m-n\right)  \mathfrak{L}_{k+m+n}\ . \label{ContractedSimplified3Virasoro}%
\end{gather}
Remarkably, the contracted 3-algebra so obtained is invariant under the
$O\left(  2\right)  $ transformation%
\begin{equation}
\mathfrak{L}_{k},\ \mathfrak{P}_{k}\mapsto\mathfrak{L}_{k}\cos\theta
+\mathfrak{P}_{k}\sin\theta,\ \mathfrak{P}_{k}\cos\theta-\mathfrak{L}_{k}%
\sin\theta\ . \label{O(2)}%
\end{equation}
An interpretation of this symmetry, as well as the validity of the FIs, is
obvious in the contracted algebra's realization as a classical 3-bracket
algebra. \ That is, \
\begin{align}
\left\{  xe^{kz},xe^{mz},xe^{nz}\right\}   &  =\left\{  ye^{kz},ye^{mz}%
,ye^{nz}\right\}  =0\ ,\nonumber\\
\left\{  xe^{kz},ye^{mz},ye^{nz}\right\}   &  =\left(  n-m\right)  ye^{\left(
k+m+n\right)  z}\ ,\nonumber\\
\left\{  ye^{kz},xe^{mz},xe^{nz}\right\}   &  =\left(  m-n\right)  xe^{\left(
k+m+n\right)  z}\ . \label{ContractedVWTernaryRealForm}%
\end{align}
In this realization the $O\left(  2\right)  $ symmetry is nothing but a
rotation about the $z$-axis. \ 

The results in (\ref{Simplified3Virasoro})-(\ref{ContractedVWTernaryRealForm}%
)\ provide the whole story, so far as we know it, for the ternary VW algebra.
\ However, for completeness, we also wish to make contact with various other
results in \cite{CFZ08}. \ By redefinition of the charges of the original
basis, it was observed in \cite{CFZ08}\ that a \textquotedblleft classical
limit\textquotedblright\ could be constructed, in which the $sl\left(  2,%
\mathbb{R}
\right)  $ Casimir went to infinity, in such a way that all FIs were OK. \ In
fact, this also just amounts to a contraction of the ternary algebra.
\ Rescaling
\begin{equation}
Q_{k}\equiv\frac{1}{\sqrt[4]{\beta\left(  1-\beta\right)  }}L_{k}%
\ ,\ \ \ R_{k}\equiv\sqrt[4]{\beta\left(  1-\beta\right)  }M_{k}\ ,
\end{equation}
substituting into (\ref{3LVW}) and (\ref{RestOf3VW}) above, and taking the
limit $\beta\rightarrow\infty$, the resulting algebra is
\begin{gather}
\left[  Q_{k},Q_{m},Q_{n}\right]  =\left(  k-m\right)  \left(  m-n\right)
\left(  k-n\right)  R_{k+m+n}\ ,\nonumber\\
\left.  \left[  R_{k},Q_{m},Q_{n}\right]  \right\vert _{\beta\rightarrow
\infty}=\left(  m-n\right)  \left(  Q_{k+m+n}+2ikR_{k+m+n}\right)
,\nonumber\\
\left[  Q_{k},R_{m},R_{n}\right]  =\left(  n-m\right)  R_{k+m+n}\ ,\nonumber\\
\left[  R_{m},R_{n},R_{k}\right]  =0\ . \label{OriginalContraction}%
\end{gather}
For finite $\beta$, there would be an additional $R_{k+m+n}$\ term in the
second relation.

Again, the contracted algebra obeys the FIs in all cases. \ This also follows
immediately from the fact that we may realize (\ref{OriginalContraction}) in
terms of classical 3-brackets. \ Explicitly we find%
\begin{gather}
\left\{  \left(  x-iky\right)  e^{kz},\left(  x-imy\right)  e^{mz},\left(
x-iny\right)  e^{nz}\right\}  =\left(  k-m\right)  \left(  m-n\right)  \left(
k-n\right)  ye^{z\left(  k+m+n\right)  }\ ,\nonumber\\
\left\{  ye^{kz},\left(  x-imy\right)  e^{mz},\left(  x-iny\right)
e^{nz}\right\}  =\left(  m-n\right)  \left(  \left(  x-i\left(  k+m+n\right)
y\right)  e^{z\left(  k+m+n\right)  }\ +2ikye^{z\left(  k+m+n\right)
}\right)  ,\nonumber\\
\left\{  \left(  x-iky\right)  e^{kz},ye^{mz},ye^{nz}\right\}  =\left(
n-m\right)  ye^{z\left(  k+m+n\right)  }\ ,\nonumber\\
\left\{  ye^{kz},ye^{mz},ye^{nz}\right\}  =0\ .
\label{ContractedVWTernaryComplexForm}%
\end{gather}
In the next section of the paper, we will explain in detail how
(\ref{ContractedVWTernaryRealForm}) and (\ref{ContractedVWTernaryComplexForm})
were first found. \ It is not difficult to guess one form given the other. \ \ 

But suppose we just transform back to the original linear combinations to
recover the classical versions of the $L$s. \ What is the effect on the
algebra? \ To answer this, let%
\begin{equation}
\ell_{n}\equiv\left(  x-\left(  \gamma+n\beta\right)  y\right)  e^{nz}%
\ ,\ \ \ p_{n}=ye^{nz}%
\end{equation}
We obtain%
\begin{align}
\left\{  \ell_{k},\ell_{m},\ell_{n}\right\}   &  =-\beta^{2}\left(
k-m\right)  \left(  k-n\right)  \left(  m-n\right)  p_{k+m+n}\ ,\nonumber\\
\left\{  \ell_{k},\ell_{m},p_{n}\right\}   &  =\left(  k-m\right)  \left(
\ell_{k+m+n}-2\beta np_{k+m+n}\right)  \ ,\nonumber\\
\left\{  \ell_{k},p_{m},p_{n}\right\}   &  =\left(  n-m\right)  p_{k+m+n}%
\ ,\nonumber\\
\left\{  p_{k},p_{m},p_{n}\right\}   &  =0\ .
\end{align}
This differs from the original, uncontracted quantal\ algebra (\ref{3LVW}) and
(\ref{RestOf3VW}) only in the $\beta$-dependent coefficients on the RHS.
\ Namely, $-\beta^{2}$ appears instead of $\beta\left(  1-\beta\right)  $ and
$-2\beta$ instead of $1-2\beta$. \ So, to repeat the observation made in
\cite{CFZ08}, we may again identify this classical 3-algebra with the infinite
$sl\left(  2,%
\mathbb{R}
\right)  $ Casimir limit, $\beta\rightarrow\pm\infty$, of the quantal algebra.

Finally, for emphasis, since the classical 3-bracket always obeys FIs, it
follows that these conditions must necessarily hold\ true for each of the
various forms of the contracted VW algebra given here.

\subsection{Classical 3-bracket algebra for exponentials}

Consider the infinite set of exponentials,%
\begin{equation}
E_{a}=\exp\left(  a\cdot r\right)  \ ,
\end{equation}
and compute the classical bracket,%
\begin{equation}
\left\{  E_{a},E_{b},E_{c}\right\}  =a\cdot\left(  b\times c\right)
~E_{a+b+c}\ . \label{Exponential3Algebra}%
\end{equation}
The indices here are 3-vectors, with $\cdot$\ and $\times$\ the usual dot and
cross products. \ This infinite algebra \emph{does} satisfy Filippov's
condition, since all classical brackets do, as well as the Bremner identity.

It is \emph{not }known how to realize (\ref{Exponential3Algebra}) as operator
3-brackets. \ Although, there is a quantal \emph{4-bracket} which gives this
3-bracket as a classical limit \cite{CZ2003}. \ To see this, compute the
operator 4-bracket $\left[  \exp\left(  a\cdot r\right)  ,\exp\left(  b\cdot
r\right)  ,\exp\left(  c\cdot r\right)  ,w\right]  $ where we assume the
exponentials do not involve $w$, and where we take $w$ and $x$, and also $y$
and $z$, to be independent canonically conjugate pairs of variables, i.e.
$\left[  w,x\right]  =i\hbar$, $\left[  y,z\right]  =i\hbar$, but $\left[
w,y\right]  =0$, etc. \ The result for the 4-bracket is then given directly by
the commutator resolution \cite{CZ,Z}.
\begin{multline}
\left[  e^{a\cdot r},e^{b\cdot r},e^{c\cdot r},w\right]  =4\hbar e^{\left(
a+b+c\right)  \cdot r}\left(  _{\ }a_{x}\sin\left(  \tfrac{1}{2}\hbar
b_{\perp}\times c_{\perp}\right)  \cos\left(  \tfrac{1}{2}\hbar\left(
b_{\perp}+c_{\perp}\right)  \times a_{\perp}\right)  \right. \\
\left.  _{\ }+b_{x}\sin\left(  \tfrac{1}{2}\hbar c_{\perp}\times a_{\perp
}\right)  \cos\left(  \tfrac{1}{2}\hbar\left(  c_{\perp}+a_{\perp}\right)
\times b_{\perp}\right)  +c_{x}\sin\left(  \tfrac{1}{2}\hbar a_{\perp}\times
b_{\perp}\right)  \cos\left(  \tfrac{1}{2}\hbar\left(  a_{\perp}+b_{\perp
}\right)  \times c_{\perp}\right)  \right)  \ , \label{Quantal4Bracket}%
\end{multline}
where $a=\left(  a_{x},a_{y},a_{z}\right)  $, $a_{\perp}=\left(  a_{y}%
,a_{z}\right)  $, $a_{\perp}\times b_{\perp}=a_{y}b_{z}-b_{y}a_{z}$, etc. \ In
the limit $\hbar\rightarrow0$, this gives the anticipated classical
3-bracket,
\begin{equation}
\frac{1}{2\hbar^{2}}\left[  \exp\left(  a\cdot r\right)  ,\exp\left(  b\cdot
r\right)  ,\exp\left(  c\cdot r\right)  ,w\right]  =a\cdot\left(  b\times
c\right)  ~\exp\left(  a+b+c\right)  \cdot r\ +O\left(  \hbar^{2}\right)  \ .
\end{equation}
Before the classical limit is taken, however, (\ref{Quantal4Bracket}) does
\emph{not} satisfy the FI: \ There are violations at $O\left(  \hbar
^{6}\right)  $ and beyond.

We now describe in more detail the relation between (\ref{Exponential3Algebra}%
) and the classical realization of the VW ternary algebra
(\ref{ContractedSimplified3Virasoro}). \ In fact, the classical VW is a
subalgebra of (\ref{Exponential3Algebra}). \ 

This may be understood as follows. \ Clearly, from (\ref{Exponential3Algebra}%
), any three exponentials with co-planar vectors will have a vanishing
classical bracket. \ By representing all the $\mathfrak{L}$s with a set of
such co-planar exponentials, and all the $\mathfrak{P}$s with another set of
co-planar exponentials, the first two lines of
(\ref{ContractedSimplified3Virasoro}) will be satisfied. \ In general then,
there are two distinct planes: \ One for the $\mathfrak{L}$s\ and one for the
$\mathfrak{P}$s. \ The remaining challenge, viewed geometrically, is to put
these two distinct planes together so that the last two lines of
(\ref{ContractedSimplified3Virasoro}) will also be satisfied. \ An obvious
guess is that the two planes should intersect at right angles. \ Another,
related guess is that the index appearing in
(\ref{ContractedSimplified3Virasoro})\ should correspond to modes along the
line of intersection of the two planes. \ 

Therefore, to play the role of the classical\ $\mathfrak{L}$s, take
$\mathfrak{l}_{k}\equiv E_{\widehat{x}+k\widehat{z}}=\exp\left(  x+kz\right)
$, while for the $\mathfrak{P}_{k}$s take $\mathfrak{p}_{k}\equiv
E_{\widehat{y}+k\widehat{z}}=\exp\left(  y+kz\right)  $. \ Some elementary
algebra then gives $\left(  \widehat{x}+m\widehat{z}\right)  \cdot\left(
\left(  \widehat{y}+n\widehat{z}\right)  \times\left(  \widehat{y}%
+k\widehat{z}\right)  \right)  =k-n$ and $\left(  \widehat{y}+k\widehat
{z}\right)  \cdot\left(  \left(  \widehat{x}+m\widehat{z}\right)
\times\left(  \widehat{x}+n\widehat{z}\right)  \right)  =m-n$ as well as
$\left(  \widehat{x}+m\widehat{z}\right)  +\left(  \widehat{y}+n\widehat
{z}\right)  +\left(  \widehat{y}+k\widehat{z}\right)  =\left(  \widehat
{x}+\widehat{y}\right)  +\widehat{y}+\left(  k+m+n\right)  \widehat{z}$ and
$\left(  \widehat{y}+k\widehat{z}\right)  +\left(  \widehat{x}+m\widehat
{z}\right)  +\left(  \widehat{x}+n\widehat{z}\right)  =\left(  \widehat
{x}+\widehat{y}\right)  +\widehat{x}+\left(  k+m+n\right)  \widehat{z}$. \ So,
modulo the common spurious vector $\left(  \widehat{x}+\widehat{y}\right)  $
we have just what we need to obtain the contracted algebra from the classical
brackets (\ref{Exponential3Algebra}). \ Now, if we incorporate the inverse of
this spurious term into the definition of a modified classical 3-bracket, as a
multiplicative factor,
\begin{equation}
\left\{  A,B,C\right\}  _{\operatorname{mod}}\equiv\frac{\partial\left(
A,B,C\right)  }{\partial\left(  x,y,z\right)  }~e^{-\left(  \widehat
{x}+\widehat{y}\right)  \cdot\overrightarrow{r}}=\frac{\partial\left(
A,B,C\right)  }{\partial\left(  x,y,z\right)  }~e^{-x-y}\ ,
\end{equation}
then we have realized on exponentials the classical, contracted VW
3-algebra.\
\begin{align}
\left\{  \mathfrak{l}_{k},\mathfrak{l}_{m},\mathfrak{l}_{n}\right\}
_{\operatorname{mod}}  &  =0\ ,\ \ \ \left\{  \mathfrak{p}_{k},\mathfrak{p}%
_{m},\mathfrak{p}_{n}\right\}  _{\operatorname{mod}}=0\ ,\nonumber\\
\left\{  \mathfrak{l}_{k},\mathfrak{p}_{m},\mathfrak{p}_{n}\right\}
_{\operatorname{mod}}  &  =\left(  n-m\right)  \mathfrak{p}_{k+m+n}%
\ ,\ \ \ \left\{  \mathfrak{p}_{k},\mathfrak{l}_{m},\mathfrak{l}_{n}\right\}
_{\operatorname{mod}}=\left(  m-n\right)  \mathfrak{l}_{k+m+n}\ .
\end{align}
But what effect does the multiplicative factor have on FIs?

It cannot obviate the FIs, because we have already verified them for the
contracted algebra. \ Another way to see this is to note the multiplicative
factor is just the Jacobian for the variable change $\left(  x,y,z\right)
\mapsto\left(  e^{x},e^{y},z\right)  $. \ In terms of these new exponential
variables the realization is%
\begin{equation}
\mathfrak{l}_{k}=xe^{kz}\ ,\ \ \ \mathfrak{p}_{k}=ye^{kz}\ ,
\label{ClassicalRealization}%
\end{equation}
where these are to be acted on by \emph{un}modified classical 3-brackets for
the new $x,y,z$ variables. Thus we obtain (\ref{ContractedVWTernaryRealForm})
of the previous section. \ 

We may summarize either (\ref{ContractedVWTernaryRealForm})\ or
(\ref{ContractedVWTernaryComplexForm}) as simply the closure of functions of
the form $xf\left(  z\right)  $ and $yg\left(  z\right)  $ under classical
3-brackets. \ A complementary algebra is given by the closure of the classical
brackets for functions of the form $\sqrt{z}f\left(  x,y\right)  $. \ This may
be expressed as a two-parameter algebra \cite{Chakrabortty} if we choose
$f\left(  x,y\right)  =\exp\left(  ax+by\right)  $ \ Again, the FI is
guaranteed to hold since only classical brackets are involved.

\section{Conclusions}

We have discussed ternary algebras and the Bremner identities which they
always obey, as well as the Filippov conditions which hold identically for
classical and ALMY 3-brackets but not for quantal brackets. \ We stress that
the Bremner identities are universal for ternary algebras, and would constrain
inclusion of central charges and related extensions in such algebras, while
the FI conditions would provide further constraints but only in more
specialized situations. \ Nevertheless, we recognize the FI is simpler, when
it applies, and often useful in specific applications. \ 

For Nambu's $su(2)$, as well as various infinite dimensional algebras, we have
provided classical realizations, which, ipso facto, ensure FI compliance. \ We
suspect that all ternary algebras based on ALMY brackets can be realized as
classical 3-brackets as well, but we have not shown this. \ We also suspect
the role of the BI in CFT operator product expansions, including
supersymmetric extensions, may be very interesting. \ We believe these open
questions are worthy of further investigation.

\begin{acknowledgments}
We record our gratitude to J Nuyts for communicating his 3-on-3-on-3 work to
us prior to publication, and for alerting us to Bremner's identity. \ This
work was supported by NSF Award 0555603, and by the US Department of Energy,
Division of High Energy Physics, Contract DE-AC02-06CH11357.
\end{acknowledgments}

\appendix

\section{Classical brackets and manifolds}

We give here a brief overview of classical brackets defined on $n$-dimensional
manifolds. \ Some features of the quantal brackets are foreshadowed in this
classical context. \ For systematic discussions, as well as guides to the
literature, see \cite{deA,Gautheron,T,Vaisman}.

We consider a general \emph{Poisson bracket} involving an antisymmetric, but
otherwise arbitrary, 2-tensor $\omega^{ab}$.%
\begin{equation}
\left\{  A,B\right\}  =\omega^{ab}~\partial_{a}A~\partial_{b}B\ .
\end{equation}
Repeated indices are implicitly summed from $1$ to $n$. \ For any $\omega
^{ab}$ this is obviously a \emph{derivation}: $\ \left\{  A,BC\right\}
=\left\{  A,B\right\}  C+B\left\{  A,C\right\}  $. \ But it is more
interesting for physics purposes that there are situations where the Poisson
bracket realizes a \emph{Lie algebra}. \ This is evident if we bracket
$\left\{  A,B\right\}  $ with $C$ to obtain two functionally independent
terms. \
\begin{equation}
\left\{  C,\left\{  A,B\right\}  \right\}  =\omega^{cd}\omega^{ab}\partial
_{d}\left(  \partial_{c}C~\partial_{a}A~\partial_{b}B\right)  +\left(
\partial_{c}C~\partial_{a}A~\partial_{b}B\right)  \omega^{cd}\partial
_{d}\omega^{ab}\ .
\end{equation}
The combination that constitutes the Lie-algebra mandated Jacobi
\textquotedblleft identity\textquotedblright\ similarly gives two terms.
\begin{equation}
\left\{  C,\left\{  A,B\right\}  \right\}  -\left\{  \left\{  C,A\right\}
,B\right\}  -\left\{  A,\left\{  C,B\right\}  \right\}  =\Omega^{abcd}%
\partial_{d}\left(  \partial_{a}A~\partial_{b}B~\partial_{c}C\right)
+\Omega^{abc}\left(  \partial_{a}A~\partial_{b}B~\partial_{c}C\right)  \ ,
\label{Jacobi}%
\end{equation}
where we have defined 4- and 3-tensors%
\begin{equation}
\Omega^{abcd}\equiv\omega^{cd}\omega^{ab}-\omega^{ac}\omega^{bd}+\omega
^{ad}\omega^{bc}\ ,\ \ \ \Omega^{abc}\equiv\omega^{cd}\partial_{d}\omega
^{ab}-\omega^{bd}\partial_{d}\omega^{ac}+\omega^{ad}\partial_{d}\omega^{bc}\ .
\end{equation}
Now the first term on the RHS of (\ref{Jacobi}) always vanishes by symmetry:
$\ $For any $\omega^{ab}=-\omega^{ba}$ the corresponding $\Omega^{abcd}$ is a
totally antisymmetric 4-tensor, and hence $\Omega^{abcd}\partial_{d}\left(
\partial_{a}A~\partial_{b}B~\partial_{c}C\right)  $ is identically zero. \ So,
for constant $\omega^{ab}$ the Jacobi identity $\left\{  C,\left\{
A,B\right\}  \right\}  -\left\{  \left\{  C,A\right\}  ,B\right\}  -\left\{
A,\left\{  C,B\right\}  \right\}  =0$ is indeed an identity for Poisson
brackets. \ 

But in general, the second term on the RHS of (\ref{Jacobi}) does not vanish
for non-constant 3-tensors. \ Hence there is a condition for the Jacobi
identity to be satisfied: \ $\Omega^{abc}=0$. \ When true, we are dealing with
a Lie algebra on a \emph{Poisson manifold}. \ If in addition $n$ is even and
the 2-tensor has an inverse, such that $\omega_{ab}\omega^{bc}=\delta
_{a}^{\ c}$, then we have a \emph{symplectic manifold}, and we can construct
the 2-form $\boldsymbol{\omega}=\omega_{ab}~dx^{a}\wedge dx^{b}$. \ In this
case the condition for the Jacobi identity to hold is easily rendered to be
$0=\partial_{a}\omega_{bc}+\partial_{b}\omega_{ca}+\partial_{c}\omega_{ab}$,
or equivalently\ just that the 2-form is closed: $\ \boldsymbol{d\omega}=0$.
\ While this is at first sight a generalization from the constant 3-tensor
case, this is somewhat illusory. \ For such closed 2-forms Darboux proved the
existence of local coordinates on the manifold such that $\omega_{ab}$ is constant.

Next, we consider a general classical \emph{Nambu 3-bracket} involving an
arbitrary, totally antisymmetric 3-tensor $\omega^{abc}$.
\begin{equation}
\left\{  A,B,C\right\}  =\omega^{abc}~\partial_{a}A~\partial_{b}B~\partial
_{c}C\ . \label{3CNB}%
\end{equation}
With this structure we encounter a few similarities with the Poisson bracket
case, but more importantly, we also encounter some dramatic differences.
\ Just like the Poisson bracket above, this is a derivation: $\ \left\{
A,B,CD\right\}  =\left\{  A,B,C\right\}  D+C\left\{  A,B,D\right\}  $. \ Also
like the Poisson bracket case, the action of one 3-bracket on another produces
two independent terms.%
\begin{equation}
\left\{  \left\{  A,B,C\right\}  ,D,E\right\}  =\omega^{abc}\omega
^{def}~\partial_{f}\left(  \partial_{a}A~\partial_{b}B~\partial_{c}%
C~\partial_{d}D~\partial_{e}E\right)  +\left(  \partial_{a}A~\partial
_{b}B~\partial_{c}C~\partial_{d}D~\partial_{e}E\right)  \omega^{def}%
\partial_{f}\omega^{abc}\ . \label{3on3}%
\end{equation}
But here the differences arise. \ What is the appropriate analogue of the
Jacobi identity? \ 

In general, \emph{there is no perfect analogue of the Jacobi identity}
involving the action of one 3-bracket on another, or a linear combination of
such. \ To see this we need only consider the case of constant $\omega^{abc}$
and make use of some elementary group theory: \ The symmetrized product of two
antisymmetric 3-tensors does \emph{not} contain a totally antisymmetric
6-tensor. \ Indeed, with the standard partition labeling of symmetric group
representations, where the sequence of integers represents the number of boxes
in the rows of a Young frame, we have
\begin{equation}
\left\{  1,1,1\right\}  _{symmetric}^{2}=\left\{  2,1,1,1,1\right\}  +\left\{
2,2,2\right\}  \ . \label{3x3symmetric}%
\end{equation}
The antisymmetric 6-tensor is found instead in the \emph{anti}symmetrized
product, $\left\{  1,1,1\right\}  _{antisymmetric}^{2}=\left\{
1,1,1,1,1,1\right\}  +\left\{  2,2,1,1\right\}  $. \ Now, because it is the
symmetrized tensor product of two $\omega$'s that appears in (\ref{3on3}),
these group properties imply that any linear combination, obtained by
permuting the entries in one classical 3-bracket acting on another,
\emph{cannot possibly vanish} without imposing some condition on $\omega
^{abc}$, and/or on the number of variables $n$. \ Neither partial nor full
antisymmetrizations of $A,B,C,D,E$ in (\ref{3on3})\ can avoid both of the
tensors on the RHS of (\ref{3x3symmetric}), and in general $\omega_{\left\{
2,1,1,1,1\right\}  }^{abcdef}\partial_{f}\left(  \partial_{a}A~\partial
_{b}B~\partial_{c}C~\partial_{d}D~\partial_{e}E\right)  \neq0\neq
\omega_{\left\{  2,2,2\right\}  }^{abcdef}\partial_{f}\left(  \partial
_{a}A~\partial_{b}B~\partial_{c}C~\partial_{d}D~\partial_{e}E\right)  $. \ We
emphasize that this is \emph{different }from the Poisson bracket case, where
the group theory is $\left\{  1,1\right\}  _{symmetric}^{2}=\left\{
1,1,1,1\right\}  +\left\{  2,2\right\}  $ and $\left\{  1,1\right\}
_{antisymmetric}^{2}=\left\{  2,1,1,1\right\}  $, and where the linear
combination of brackets in the Jacobi identity\ serves to single out $\left\{
1,1,1,1\right\}  $ and thus eliminate the $\omega\omega$ term for \emph{all}
antisymmetric 2-tensors.

Admittedly, there is one \emph{very special} case where group theory does not
impose an impasse for classical 3-on-3-bracket identities, namely,\ $n=3$.
\ When there are only three independent variables, the $\left\{
2,1,1,1,1\right\}  $ representation is absent! \ In this special case we
obtain the FI.%
\begin{equation}
\left\{  \left\{  A,B,C\right\}  ,D,E\right\}  =\left\{  \left\{
A,D,E\right\}  ,B,C\right\}  +\left\{  A,\left\{  B,D,E\right\}  ,C\right\}
+\left\{  A,B,\left\{  C,D,E\right\}  \right\}  \ . \label{Filippov}%
\end{equation}
Of course, for this case there is only one antisymmetric 3-tensor, namely,
that of Kronecker, so $\omega^{abc}\propto\varepsilon^{abc}$. \ But, alas,
Filippov's condition for 3-brackets does not go willingly into higher
dimensional manifolds.

More generally, if we define a classical N-bracket as $\left\{  A_{1}%
,A_{2},\cdots,A_{N}\right\}  =\omega^{a_{1}a_{2}\cdots a_{N}}\partial_{a_{1}%
}A_{1}\partial_{a_{2}}\cdots\partial_{a_{N}}A_{N}$, then the even N cases are
like that of the Poisson bracket, while the odd N cases are like that for the
3-bracket. \ When N is even, we have $\left\{  1^{2N}\right\}  \subset\left\{
1^{N}\right\}  _{symmetric}^{2}$, so for general constant $\omega$'s the
action of one N-bracket on another vanishes \emph{when totally antisymmetrized
over the 2N-1 entries in the double bracket}. \ When N is odd, we have
$\left\{  1^{2N}\right\}  \subset\left\{  1^{N}\right\}  _{antisymmetric}^{2}%
$, so for constant $\omega$'s the action of any permuted linear combination of
one N-bracket on another does \emph{not} vanish without imposing some
condition on the $\omega$'s and/or on the number of variables. \ For example,
if there are only N variables and $\omega$ is the Kronecker tensor, the
N-bracket version of the FI holds. \


\newpage

\begin{thebibliography}{99}                                                                                               %


\bibitem {Ahn}C Ahn, \textquotedblleft Towards Holographic Gravity Dual of N=1
Superconformal Chern-Simons Gauge Theory\textquotedblright\
JHEP \textbf{0807}, 101 (2008) [arXiv:0806.4807 [hep-th]].


\bibitem {A}H Awata, M Li, D Minic, and T Yoneya, \textquotedblleft On the
Quantization of Nambu Brackets,\textquotedblright\ JHEP 0102 (2001) 013 arXiv:hep-th/9906248.



\bibitem {Axenides}M Axenides and E Floratos, \textquotedblleft Nambu-Lie
3-Algebras on Fuzzy 3-Manifolds\textquotedblright\
JHEP \textbf{0902}, 039 (2009) [arXiv:0809.3493 [hep-th]].


\bibitem {B}J Bagger and N Lambert, \textquotedblleft Gauge Symmetry and
Supersymmetry of Multiple M2-Branes,\textquotedblright\ Phys. Rev.
\textbf{D77}, 065008 (2008) [arXiv:0711.0955 [hep-th]]; \ J Bagger and N
Lambert, \textquotedblleft Comments On Multiple M2-branes,\textquotedblright%
\ JHEP 0802, 105 (2008) [arXiv:0712.3738 [hep-th]].

\bibitem {S}M A Bandres, A E Lipstein, and J H Schwarz, \textquotedblleft N =
8 Superconformal Chern--Simons Theories\textquotedblright\ [arXiv:0803.3242v4
[hep-th]]; \ Bandres, A E Lipstein, and J H Schwarz, \textquotedblleft
Ghost-Free Superconformal Action for Multiple M2-Branes\textquotedblright%
\ [arXiv:0806.0054v1 [hep-th]].

\bibitem {Bonelli}G Bonelli, A Tanzini and M Zabzine, \textquotedblleft
Topological branes, p-algebras and generalized Nahm
equations\textquotedblright\ Phys. Lett. B \textbf{672}, 390(2009)
[arXiv:0807.5113 [hep-th]].


\bibitem {Bremner1995}M R Bremner, \textquotedblleft Varieties of
anticommutative n-ary algebras\textquotedblright\ J. Algebra \textbf{191}
(1997) 76-88.

\bibitem {Bremner}M R Bremner, \textquotedblleft Identities for the ternary
commutator\textquotedblright\ J. Algebra \textbf{206} (1998) 615--623; \ M R
Bremner and L A Peresi, \textquotedblleft Ternary analogues of Lie and Malcev
algebras\textquotedblright\ Linear Algebra and its Applications \textbf{414}
(2006) 1-18.

\bibitem {Chakrabortty}S Chakrabortty, A Kumar and S Jain, \textquotedblleft
w$_{\infty}$ 3-algebra\textquotedblright\
JHEP \textbf{0809}, 091 (2008) [arXiv:0807.0284 [hep-th]].


\bibitem {Chu}C S Chu, P M Ho, Y Matsuo, and S Shiba, \textquotedblleft
Truncated Nambu-Poisson Bracket and Entropy Formula for Multiple
Membranes\textquotedblright\
JHEP \textbf{0808}, 076 (2008) [arXiv:0807.0812 [hep-th]].


\bibitem {CFZ08}T L Curtright, D B Fairlie, and C K Zachos, \textquotedblleft
Ternary Virasoro-Witt Algebra\textquotedblright\ Phys. Lett. B \textbf{666},
386-390 (2008) [arXiv:0806.3515 [hep-th]].

\bibitem {CFZ08Misstatement}Rather, Filippov was following up on earlier
studies that had appeared in the mathematics literature, primarily by Kurosh
\cite{Kurosh}. \ This corrects an erroneous remark on the history of
Filippov's work, as made in \cite{CFZ08}.

\bibitem {CZ2002}T L Curtright and C K Zachos, \textquotedblleft Deformation
quantization of superintegrable systems and Nambu mechanics\textquotedblright%
\ New J.\ Phys.\ \textbf{4} (2002) 83 [arXiv:hep-th/0205063].


\bibitem {CZ}T L Curtright and C K Zachos, \textquotedblleft Classical and
quantum Nambu mechanics\textquotedblright\ Phys. Rev. \textbf{D68}, 085001
(2003) [hep-th/0212267].

\bibitem {CZ2003}T L Curtright and C K Zachos, \textquotedblleft Branes,
strings, and odd quantum Nambu brackets\textquotedblright\ in Quantum Theory
and Symmetries: Proceedings of the 3rd International Symposium, Cincinnati,
10-14 September, 2003, P Argyres et al (eds), World Scientific Publishers,
2004, pp 206-217 [hep-th/0312048].

\bibitem {deA}J A de Azcarraga, J M Izquierdo, and J C Perez Bueno,
\textquotedblleft An introduction to some novel applications of Lie algebra
cohomology and physics\textquotedblright\ Rev. R. Acad. Cien. Exactas Fis.
Nat. Ser. A Mat. \textbf{95} (2001) 225-248 [physics/9803046];
\textquotedblleft On the generalizations of Poisson
structures\textquotedblright\ J. Phys. A \textbf{30} (1997) L607-L616 [hep-th/9703019].

\bibitem {F}V T Filippov, \textquotedblleft n-Lie Algebras\textquotedblright%
\ Sib. Mat. Zh. \textbf{26} (1985) 126-140 (English translation: Sib. Math.
Journal \textbf{26} (1986) 879-891).

\bibitem {Garousi}M R Garousi, A Ghodsi, and M Khosravi, \textquotedblleft On
thermodynamics of N=6 superconformal Chern-Simons theories at strong
coupling\textquotedblright\
JHEP \textbf{0808}, 067 (2008) [arXiv:0807.1478 [hep-th]].


\bibitem {Gautheron}P Gautheron, \textquotedblleft Some remarks concerning
Nambu mechanics\textquotedblright\ Lett. Math. Phys., V 37, (1996) 103-116.
\ DOI 10.1007/BF00400143

\bibitem {GT}P Goddard and C B Thorn, \textquotedblleft Compatibility of the
Dual Pomeron with Unitarity and the Absence of Ghosts in the Dual Resonance
Model\textquotedblright\ Phys. Lett. B \textbf{40} (1972) 235-238.

\bibitem {G}J Gomis, G Milanesi, J G Russo, \textquotedblleft Bagger-Lambert
Theory for General Lie Algebras\textquotedblright\ [arXiv:0805.1012v2
[hep-th]]; \ J Gomis, D Rodriguez-Gomez, M Van Raamsdonk, and H Verlinde,
\textquotedblleft The Superconformal Gauge Theory on
M2-Branes\textquotedblright\ [arXiv:0806.0738v1 [hep-th]].

\bibitem {Gustavsson}A Gustavsson, \textquotedblleft Algebraic structures on
parallel M2-branes\textquotedblright\ [arXiv:0709.1260[hep-th]]; \ A
Gustavsson, \textquotedblleft Selfdual strings and loop space Nahm
equations\textquotedblright\ [arXiv:0802.3456[hep-th]].

\bibitem {Hanaki}K Hanaki and H Lin, \textquotedblleft M2-M5 Systems in N=6
Chern-Simons Theory\textquotedblright\
JHEP \textbf{0809}, 067 (2008) [arXiv:0807.2074 [hep-th]].


\bibitem {H}P-M Ho, R-C Hou, and Y Matsuo, \textquotedblleft Lie 3-Algebra and
Multiple M2-branes\textquotedblright\ [arXiv:0804.2110v2 [hep-th]]; \ P-M Ho,
Y Imamura, and Y Matsuo, \textquotedblleft M2 to D2
revisited\textquotedblright\ [arXiv:0805.1202v2 [hep-th]]; \ P-M Ho and Y
Matsuo \textquotedblleft M5 from M2\textquotedblright\ [arXiv:0804.3629v2 [hep-th]].

\bibitem {Hoppe}J Hoppe, \textquotedblleft On M-Algebras, the Quantisation of
Nambu-Mechanics, and Volume Preserving Diffeomorphisms\textquotedblright%
\ Helv. Phys. Acta \textbf{70} (1997) 302-317 [hep-th/9602020].

\bibitem {IW}E In\"{o}n\"{u} and E P Wigner, \textquotedblleft On the
contraction of groups and their representations\textquotedblright\ Proc. Nat.
Acad. Sci. U.S.A., 1953, V.39, 510--524.

\bibitem {Kurosh}A G Kurosh, \textquotedblleft Free sums of multioperator
algebras\textquotedblright\ Sib. Mat. Zh. \textbf{1} (1960) 62-70;
\textquotedblleft Multioperator rings and algebras\textquotedblright\ Usp.
Mat. Nauk \textbf{24}, No. 1, 3-15 (1969).

\bibitem {Larsson}T A Larsson, \textquotedblleft Virasoro 3-algebra from
scalar densities\textquotedblright\
[arXiv:0806.4039 [hep-th]].


\bibitem {KM}H Lin, \textquotedblleft Kac-Moody Extensions of 3-Algebras and
M2-branes\textquotedblright\ [arXiv:0805.4003 [hep-th]].

\bibitem {N}Y Nambu, \textquotedblleft Generalized Hamiltonian
Dynamics\textquotedblright\ Phys. Rev. \textbf{D7} (1973) 2405-2412.

\bibitem {Nuyts}J Nuyts, \textit{in preparation}.

\bibitem {Sahoo}D Sahoo and M C Valsakumar, \textquotedblleft Nambu mechanics
and its quantization\textquotedblright\ Phys. Rev. \textbf{A46} (1992) 4410 - 4412.

\bibitem {T}L Takhtajan, \textquotedblleft On foundation of the generalized
Nambu mechanics\textquotedblright\ Comm. Math. Phys. \textbf{160} (1994)
295--315 [hep-th/9301111].

\bibitem {Vaisman}I Vaisman, \textquotedblleft A survey on Nambu--Poisson
brackets\textquotedblright\ Acta Math. Univ. Comenianae Vol. LXVIII, 2(1999),
pp. 213--241 [arXiv:math/9901047].

\bibitem {Z}An efficient way to carry out the computation is to use the
methods of deformation quantization. \ For example, see C K Zachos, D B
Fairlie, and T L Curtright, \textit{Quantum Mechanics in Phase Space}, World
Scientific, 2005.
\end{thebibliography}
\end{document}